\documentclass[11pt]{amsart}
\usepackage{geometry}
\usepackage{graphicx}
\usepackage{amsmath}
\usepackage{amssymb}
\usepackage{mathtools}
\usepackage{epstopdf}
\usepackage[mathscr]{euscript}

\newcommand{\al}{\alpha}
\newcommand{\pa}{\partial}
\newcommand{\ep}{\epsilon}

\newcommand{\ta}{\tau}
\newcommand{\ga}{\gamma}
\newcommand{\Om}{\Omega}
\newcommand{\om}{\omega}

\newcommand{\De}{\Delta}

\newcommand{\rar}{\rightarrow}
\newcommand{\lrar}{\leftrightarrow}

\newcommand{\non}{\nonumber}

\begin{document}
\title{$F_4$ Quantum Integrable, rational and trigonometric models: space-of-orbits view}

\author{A.V.~Turbiner and J.C.~L\'opez Vieyra}

\address{Instituto de Ciencias Nucleares, \\
Universidad Nacional Aut\'onoma de M\'exico, 04510 Mexico D.F.}
\date{November 26, 2013}
\email{turbiner@nucleares.unam.mx, vieyra@nucleares.unam.mx}

\begin{abstract}
Algebraic-rational nature of the four-dimensional, $F_4$-invariant integrable quantum  Hamiltonians, both rational and trigonometric, is revealed and reviewed. It was shown
that being written in $F_4$ Weyl invariants, polynomial and exponential, respectively, both similarity-transformed Hamiltonians are in algebraic form, they are quite similar the second order differential operators with polynomial coefficients; the flat metric in the Laplace-Beltrami operator has polynomial (in invariants) matrix elements. Their potentials are calculated for the first time: they are meromorphic (rational) functions with singularities at the boundaries of the configuration space. Ground state eigenfunctions are algebraic functions in a form of polynomials in some degrees. Both Hamiltonians preserve the same infinite flag of polynomial spaces with characteristic vector $(1, 2, 2, 3)$, it manifests exact solvability. A particular integral common for both models is derived. The first polynomial eigenfunctions are presented explicitly.
\end{abstract}

\maketitle

\section{Introduction}

In four dimensional Euclidian space there exist several remarkable completely integrable quantum systems originally discovered in the Hamiltonian reduction method (for a review, see
\cite{Olshanetsky:1983}). These are particular cases of the Calogero-Moser-Sutherland systems and all of them are characterized by a discrete symmetry
given by Weyl group $W$, acting on root spaces $A_4, BC_4, F_4$ and $H_4$.
In such models, the potential has four possible forms (rational $(\sim 1/x^2)$,
trigonometric $(\sim 1/{(\sin x)}^2)$, hyperbolic $(\sim 1/{(\sinh x)}^2)$ and elliptic $(\sim \wp(x))$).
All these $4D$ Hamiltonians are exactly solvable in terms of Weyl-invariant variables. For rational and trigonometric (also hyperbolic) systems, exact solvability was obtained in an explicit constructive fashion in \cite{Ruhl:1995,Brink:1997,blt,H4}.
The spectrum is found explicitly, in closed analytic form being a first/second degree polynomial  in the quantum numbers for the rational/trigonometric systems respectively.

The purpose of the present paper is to revisit the $F_4$-invariant model in its rational and trigonometric forms showing how it looks like in the space of orbits, in the space of $F_4$ invariants. For the rational $F_4$ model it is mainly reduced to a review the results obtained in previous studies \cite{blt} (see also \cite{BLT-rational}) and their reinterpretation. In addition, we show for the first time the explicit form of the ground state eigenfunction and the original potential in terms of Weyl invariant variables. For the $F_4$ trigonometric model we present a more succinct solution compared to our previous analysis \cite{blt} (see also \cite{BLT-trigonometric}), and we also present for the first time the explicit form of the ground state eigenfunction and the original potential in terms of trigonometric Weyl invariant variables. For both $F_4$ rational and trigonometric models the extra, particular integral
\cite{Turbiner:2013p} will be derived and it will be shown that is the same in the space-of-orbits.
Eventually, it will be given a place to both models in the space-of-orbit formalism
\cite{Turbiner:2011,Turbiner:2013}.

\bigskip

\section{Generalities}

\bigskip

The $F_4$-Weyl-invariant Calogero-Moser-Sutherland Hamiltonian is defined as follows
\begin{equation}
\label{OPHams}
 H = - \frac{1}{2}\sum_{j=1}^{N} \frac{\pa^2}{\pa x_j^2} +
 \frac{1}{2}\sum_{\al\in{\mathcal F_4}^+} g_{|\al|} {|\al|^2}\, {\mathcal V}((\al\cdot x))\ ,
\end{equation}
where ${\mathcal F_4}^+$ is the set of positive roots, $N=4$ is the rank  of the root system
(the dimension of the configuration space), and $g_{|\al|}$
are coupling constants. For roots of the same length the same coupling constant $g_{|\al|}$
is assigned.
In (\ref{OPHams})  $x$ denotes the $(N=4)$-dimensional vector $x=(x_1,x_2,x_3, x_4)$.
The potential consists of a linear superposition of the terms ${\mathcal V}((\al \cdot x))$ of the argument $(\al\cdot x)$ where $\al \in{\mathcal F_4}^+$ is a positive root (see \cite{Olshanetsky:1983}). For the rational case the four-dimensional isotropic harmonic oscillator potential is usually added to the Hamiltonian (\ref{OPHams}).

The set ${\mathcal F_4}^+$ consists of 24 positive roots:
\begin{equation}
\label{F4posroots}
\begin{array}{cc}
\al_{\rm short} =  e_1,e_2,e_3,e_4,  \, \frac{1}{2}( e_1 \pm e_2 \pm e_3 \pm e_4)\,, & |\al_{\rm short}|=1 \,,\\[12pt]
\al_{\rm long} =  e_i \pm e_j ,\quad i>j   \quad i,j=1,\ldots 4\,, & |\al_{\rm long}|=\sqrt{2}\,,
\end{array}
\end{equation}
where ${e_1, e_2, e_3, e_4}$ are the standard orthonormal  vectors in the 4-dimensional Euclidean space. From a geometrical viewpoint, the short root vectors in (\ref{F4posroots}) are, in fact, the vertices of the 4-dimensional 24-cell regular polytope (see~\cite{Polytopes}), while the long roots are the vertices of its dual 24-cell. The rotational symmetry group of the 24-cell has the order 576.  The full symmetry group of the 24-cell is the $F_4$ Weyl group (of order 1152), which is generated by reflections on the hyperplanes orthogonal to the $F_4$ roots.

\bigskip

\section{$F_4$ Rational Model}

\bigskip

In a four-dimensional Euclidean space with coordinates $x=(x_1, x_2, x_3, x_4)$ the Hamiltonian of the $F_4$ Rational Model is given by
\begin{eqnarray}
H^{\rm rat}_{F_4} &=& \frac{1}{2} \sum_{i =1}^{4}\left( - \partial^2_{x_i}   +
 \omega^2 \, x_i^2  \right)
+ g_{\ell} \, \sum_{j<i =1}^{4}
  \left[
\frac{1}{(x_i +x_j)^2} + \frac{1}{(x_i - x_j)^2}
  \right]
\label{HamF4rat} \\
  &+&
\frac{g_s}{2} \sum_{i =1}^{4} \frac{1}{x_i^2}
+ \, 2g_s \sum_{\nu{\rm 's}=0,1}  \frac{1}{
\left[
x_1 + (-)^{\nu_2}x_2 + (-)^{\nu_3}x_3 + (-)^{\nu_4}x_4
\right]^2} \,, \non
\end{eqnarray}
with $\sum \nu{\rm 's} =$ even. The parameters
$g_{\ell}$,   $g_{s}$ are coupling constants associated with {\it long} and {\it short} roots respectively, and $\omega$ is the frequency of $4D$-isotropic harmonic oscillator potential. The configuration space is the $F_4$-Weyl Chamber.

\bigskip

The ground state eigenfunction of the Hamiltonian (\ref{HamF4rat}) has the form
\begin{equation}
\label{GSF4rat}
\psi_0(x) = (\Delta_+ \, \Delta_-)^\nu (\Delta \, \Delta_0)^\mu\,
{\rm e}^{-\frac{1}{2} \omega  \sum_{i=1}^{4} x_i^2 }  \,.
\end{equation}
where
 \begin{equation}
\label{DD}
{\De_+\,\De_-} = \prod_{i<j=1}^4    (x_i \pm x_j)\,,\quad
  {\De\,\De_0} = \frac{1}{256}\prod_{i=1}^4  x_i (x_1 \pm x_2 \pm x_3 \pm x_4)\,,
\end{equation}
are generalized Vandermonde determinants.
The exponents $\nu,\mu$ of the ground state eigenfunction  (\ref{GSF4rat}) are connected to the coupling constants  $g_{\ell}$ , $g_{s}$  by the relations
\begin{equation}
\label{gg}
g_{\ell} = \nu(\nu-1), \,\,  g_{s} = \mu(\mu-1)\,,
\end{equation}
and the ground state energy is
\begin{equation}
E_0 = 2\omega ( 1 + 6\mu + 6\nu)\,.
\end{equation}
The eigenfunctions  $\psi_n(x)$  of the Hamiltonian (\ref{HamF4rat}) have a factorized form
\begin{equation}
\label{psinF4rat}
\psi_n(x) = \psi_0(x) {\phi_n(x)}\,,
\end{equation}
where the common factor $\psi_0(x)$ is the ground state eigenfunction (\ref{GSF4rat}) and $n$ is multi-index given by four quantum numbers of the eigenstate. The functions ${\phi_n(x)}$ in (\ref{psinF4rat}) are eigenfunctions of the transformed Hamiltonian

\begin{equation}
\label{hF4rat}
\mathfrak{h}_{F_4}^{\rm rat} \equiv -2 {\psi_0}^{-1} \, { \left[ H_{F_4}^{({\rm rat})} -E_0 \right] }\,\,{\psi_0}  \,,
\end{equation}

Then, the eigenvalue problem becomes
\[
  \mathfrak{h}_{F_4}^{\rm rat}\,  \phi_n = \ep_n \phi_n\,,
\]
where $\epsilon_n =  -2(E_n-E_0)$.

The Hamiltonian (\ref{hF4rat}) acquires an algebraic form
({\it i.e.} the form of a second order differential operator with polynomial coefficients) when it
is written in terms of  an algebraically independent set of Weyl invariant variables defined as
\begin{equation}
\label{orbitvarsF4rat}
t_a ^{(\Om)}= \frac{1}{12} \sum_{k=1}^{|\Omega|}  (\alpha_k,x)^a, \qquad \alpha_k \in \Omega \,,
\end{equation}
where $a = 2,6,8,12$ are the degrees of the polynomial invariants of the $F_4$ Weyl group, and $\alpha_k$ are the roots of a conveniently chosen Weyl orbit $\Om$. In the present study, \(\Om \) was chosen as the orbit generated by the
{\em highest root} \footnote{Since the root $e_1+e_2 \in \Omega(e_3 + e_4)$, the orbit actually coincides with the orbit used in the previous studies\cite{blt, BLT-rational}.} $e_3 + e_4$ with size $|\Om|=24$. As a matter of fact, this orbit consists of all
(positive and negative) long roots. The factor $1/12$ in the definition (\ref{orbitvarsF4rat}) was introduced for convenience.

The fact that the invariants  of given degrees (\ref{orbitvarsF4rat}) are defined up to  polynomials in invariants of lower degree was used in \cite{BLT-rational} to obtain a particular set of invariants of fixed degrees $\ta_a, {\scriptstyle (a= 2,6,8,12)}$ leading to a simple algebraic form of the Hamiltonian. These invariants are related to orbital ones (\ref{orbitvarsF4rat}) in the following manner
\begin{align}
 \ta_2 &= t_2^{(\Om)} \ ,   \quad  \ta_6 = \frac{1}{12} t_6^{(\Om)}-\frac{1}{12}\big(t_2^{(\Om)}\big)^3  \,, \label{tauvarsF4rat}\\
 \ta_8 &= \frac{1}{80} t_8^{(\Om)} -\frac{1}{30} t_2^{(\Om)} t_6^{(\Om)}
  + \frac{1}{48} \big(t_2^{(\Om)}\big)^4 \ ,\non\\
 \ta_{12} &=
 \frac{1}{720} t_{12}^{(\Om)} - \frac{5}{288} \big(t_2^{(\Om)}\big)^2
 t_8^{(\Om)} + \frac{1}{27} \big(t_2^{(\Om)}\big)^3 t_6^{(\Om)} -
 \frac{29}{1440} \big(t_2^{(\Om)}\big)^6 - \frac{1}{1080} \big(t_6^{(\Om)}\big)^2\ , \non
\end{align}
and are given explicitly in Appendix A. In terms of the $\ta_2, \ta_6, \ta_8, \ta_{12}$, the  Hamiltonian (\ref{hF4rat})
\begin{equation}
\label{halgF4rat}
 \mathfrak{h}_{F_4}^{\rm rat}(\tau) =
  {\mathcal A}_{\mu\, \eta} \frac{\partial^2\ }{\partial\tau_\mu \partial \tau_\eta}  +
  {\mathcal B}_\mu   \,\frac{\partial}{\partial \tau_\mu}\,,
\end{equation}
takes the algebraic form
where the coefficient functions ${\mathcal A}_{\mu,\eta}, {\mathcal B}_{\mu}$ are polynomial in $\tau_a$,
\begin{align}
\mathfrak{h}_{F_4}^{\rm rat}(\tau) = &
 4\ta_2 \frac{\pa^2}{\pa \ta_2^2}
   + \frac{2}{3} \ta_2 ({\ta_2} \ta_6+10 \ta_8)\frac{\pa^2}{\pa \ta_6^2}
   + 2(\ta_2 \ta_{12} + 2\ta_6\ta_8)\frac{\pa^2}{\pa \ta_8^2}
   + 6\ta_8(\ta_2 \ta_{12} + 2\ta_6\ta_8) \frac{\pa^2}{\pa \ta_{12}^2}
   \non\\
&\mbox{} + 24 \ta_6 \frac{\pa^2}{\pa \ta_2\pa \ta_6}
   +32\ta_8 \frac{\pa^2}{\pa \ta_2\pa \ta_8}
   +48\ta_{12}\frac{\pa^2}{\pa \ta_2\pa \ta_{12}}
   +\frac{8}{3} ({\ta_2}^2 \ta_8 + 6\ta_{12}) \frac{\pa^2}{\pa\ta_6\pa \ta_8}
\non\\
&\mbox{}  + 4({\ta_2}^2 \ta_{12} + 8{\ta_8}^2)\frac{\pa^2}{\pa \ta_6\ta_{12}}
 + 4 (2 \ta_2 {\ta_8}^2+3\ta_6\ta_{12}) \frac{\pa^2}{\pa \ta_8 \ta_{12}}
\label{halgF4ratexplicit}\\
&\mbox{} - 4[\om \ta_2 - 2(6\nu +6 \mu +1)]\frac{\pa}{\pa \ta_2}
-[12 \om \ta_6 - {\ta_2}^2 (4\nu +2 \mu + 1)]\frac{\pa}{\pa \ta_6}\non\\
&- 4[4\om {\ta_8}-{\ta_6}(1+3\nu)]\frac{\pa}{\pa \ta_8}-4[6\om
\ta_{12}-\ta_2 \ta_8 (2+3\nu)]\frac{\pa}{\pa \ta_{12}} \ . \non
\end{align}

This algebraic Hamiltonian has infinitely many invariant polynomial subspaces of the form
\begin{equation}
\label{minflagF4rat}
 {\mathcal P}_{n}^{({1},{2},{2},{3})} =
\langle  \ta_2^{p_1} \ta_6^{p_3} \ta_8^{p_4} \ta_{12}^{p_6} | \ 0
\leq {1} p_1 + {2} p_3 + {2} p_4 + {3} p_6 \leq n \rangle\,, \quad n=0,1,2,\ldots
\end{equation}
which are marked by the characteristic vector $(1,2,2,3)$ of minimal possible {\it gradings} for the invariants $\ta_2, \ta_6, \ta_8, \ta_{12} $, respectively.
Incidentally, this characteristic vector coincides with the $F_4$ highest root among short roots when written in the basis of simple roots. Interestingly, the invariant subspaces (\ref{minflagF4rat}) form an infinite flag
\begin{equation}
\label{minflag}
{\mathcal P}_0^{({1},{2},{2},{3})}  \subset  {\mathcal P}_1^{({1},{2},{2},{3})}  \subset {\mathcal P}_2^{({1},{2},{2},{3})}  \subset \ldots
 \subset  {\mathcal P}_n^{({1},{2},{2},{3})}   \subset \ldots \,,
\end{equation}
which we call {\it minimal flag}. Since the algebraic Hamiltonian $\mathfrak{h}_{F_4}^{\rm rat}(\tau)$ ($\tau$ stands for $\ta_2, \ta_6, \ta_8, \ta_{12}$)
preserves this infinite flag (\ref{minflag}), thus it is exactly solvable (for a discussion see e.g. \cite{BLT-rational}). The eigenfunctions from the spaces ${\mathcal P}_{n}^{({1},{2},{2},{3})},\ n=0,1,2$ are presented in Appendix B.

There exists an algebra of differential operators for which the space ${\mathcal P}_{n}^{(1,2,2,3)}$  is the finite-dimensional irreducible representation space \cite{blt}. This algebra is called $f^{(4)}$ . This algebra is infinite-dimensional but finitely-generated.
The rational $F_4$ Hamiltonian in algebraic form $\mathfrak{h}_{F_4}^{\rm rat}(\tau)$ can be rewritten in terms of the generators of this algebra (see \cite{blt}).

\subsection{Flat Metric of the $F_4$-rational Model.}

Gauge rotated Hamiltonian $\mathfrak{h}_{F_4}^{\rm rat}$ (\ref{hF4rat}) in algebraic form (\ref{halgF4rat}) can be written as
\begin{equation}
\label{hF4ratD}
\mathfrak{h}_{F_4}^{\rm rat}\ =\ \De_g\ +\ {\mathcal C}^b \frac{\pa}{\pa \ta_{b}}\ ,
\end{equation}
where the Laplace-Beltrami operator
\begin{equation}
\label{LBrat}
    \De_g\ =\ \frac{1}{g^{1/2}}\frac{\pa}{\pa \ta_{a}}\ (g^{a b} g^{1/2})\ \frac{\pa}{\pa \ta_{b}}\ =\ g^{a b} \frac{\pa^2}{\pa \ta_{a}\pa \ta_{b}}\ +\ g^{b} \frac{\pa}{\pa \ta_{b}}\ ,\ g^{b} \equiv \frac{1}{g^{1/2}}\frac{\pa}{\pa \ta_{a}}\ (g^{a b} g^{1/2})
\end{equation}
with $a,b=2,6,8,12$ and $g$ is determinant. The symmetric in $a \rar b$ metric, $g^{a b}=g^{b a}$ has polynomial in $\ta$ matrix elements

\begin{equation}
g^{a b}\ = \left| \begin{array}{cccc}
 4\ta_2 & 12 \ta_6 & 16 \ta_8 & 24 \ta_{12}\\
     &    \\
  & \frac{2}{3} \ta_2 (\ta_2 \ta_6+10\ta_8)
 & \frac{4}{3} ({\ta_2}^2 \ta_8 + 6\ta_{12}) & 2({\ta_2}^2 \ta_{12} + 8{\ta_8}^2)\\
     &    \\
  &  & 2(\ta_2 \ta_{12} + 2\ta_6 \ta_8) & 2(2\ta_2 {\ta_8}^2+3\ta_6\ta_{12})\\
     &    \\
   &  &
   &  6\ta_8 (\ta_2\ta_{12}+2\ta_6\ta_8)
 \end{array} \right|\ .
\end{equation}
This metric has a special property: any component $g^b$ of the vector $\vec{g}$ in (\ref{LBrat}) is also polynomial in $\ta$
\[
    \vec{g}\ =\ (8, \ta_2^2, \ta_6, 2 \ta_2 \ta_6 )\ .
\]
The same property has the vector ${\mathcal C}^b$ (see (\ref{hF4ratD})) which characterizes the interaction in (\ref{HamF4rat})
\[
     -\vec{\mathcal C}\ =\ \bigg(
     4[\om \ta_2 - 12(\nu +\mu)],\
     2[6\om \ta_6 -  (2\nu +\mu){\ta_2}^2],\
     4[4\om {\ta_8}-3 \nu {\ta_6}],\
     12[2\om\ta_{12}-\nu \ta_2 \ta_8 ]
        \bigg)
\]

\subsection{Spectrum of the $F_4$-rational Model.}

Since the invariants $\ta_2, \ta_6, \ta_8, \ta_{12}$ have a natural ordering given by their degrees, the Hamiltonian (\ref{halgF4ratexplicit}) has a triangular form
in the basis of monomials
\(
\tau_2^{n_1} \tau_6^{n_2}\tau_8^{n_3}\tau_{12}^{n_4} \
\)
The spectrum is easily obtained:
\begin{equation}
 \label{F4ratspectrum}
\epsilon_{n_1,n_2, n_3, n_4} = -4\om (2n_1 + 6n_2 + 8n_3 + 12n_4)\ ,\ n_{1,2,3,4}=0,1,2,\ldots\ .
\end{equation}
The spectrum (\ref{F4ratspectrum}) does not depend on couplings $g_{\ell}, g_s$ and is equidistant with a degeneracy corresponding to the  partitions of integer $N= 2n_1 + 6n_2 + 8n_3 + 12n_4$. So, the interactions in (\ref{HamF4rat}) have the only effect of modifying the degeneracy of the  energies corresponding to the 4-dimensional isotropic harmonic oscillator part of the Hamiltonian (\ref{HamF4rat}).

\subsection{Ground state eigenfunction and Potential in $\ta$ variables}

The ground state eigenfunction (\ref{GSF4rat}) and the original potential can be written in terms of the invariant variables (\ref{tauvarsF4rat}). After straightforward but sometimes rather tedious calculations we find that
\begin{equation*}
\big(\Delta_+\,\Delta_- \big)^{2}\ =\ 64\, (- 3\, \ta_{12}^2 + 4\, \ta_8^3) \equiv 64\ P_1(\ta) \  ,
\end{equation*}
\begin{align}
 \big(\Delta_0\,\Delta \big)^{2} = & \frac{1}{4096}
 (- 192\, \ta_{12}^2 + 256\, \ta_8^3 + 144\, \ta_6^2\, \ta_{12}
 - 27\, \ta_6^{4} - 192\, \ta_2\, \ta_6\, \ta_8^2  \non \\ & +
 48\, \ta_2^2\, \ta_8\, \ta_{12} + 30\, \ta_2^2\, \ta_6^2\, \ta_8
 - 12 \ta_2^3\, \ta_6\, \ta_{12} + \frac{1}{2} \ta_2^3\, \ta_6^3
\non \\[3pt] &
 +  \ta_2^{4}\, \ta_8^2 - \frac{1}{2} \ta_2^5\, \ta_6\, \ta_8 +
 \frac{1}{6} \ta_2^6\, \ta_{12}) \equiv \frac{1}{2^{12}}P_2(\ta)\ , \non
\end{align}
hence, the unnormalized ground state eigenfunction (\ref{GSF4rat}) is
\begin{equation}
\label{GSF4ratintau}
\psi_0(\ta)\ =\ P_1(\ta)^{\nu/2} P_2(\ta)^{\mu/2}\,
{\rm e}^{-\frac{1}{2} \om  \ta_2 }  \ ,
\end{equation}
(cf. (\ref{GSF4rat})).

The potential $ V_{F_4}^{\rm rat}$ can be represented as
\begin{equation}
\label{VF4rat}
  V_{F_4}^{\rm rat} = {\om}^{2} V_{\om} +  g_{\ell} V_\nu + g_s V_\mu\ ,
\end{equation}
where
\begin{equation}
\label{VomegaF4rat}
V_\om \ = \ \frac{{\tau_2}}{2}  \,,
\end{equation}
\begin{equation}
\label{VnuF4rat}
V_\nu \ =\ -\ 9 \,{\frac {\left( {\tau_2}\,{\tau_{12}}-2\,{\tau_6}\,{\tau_8} \right)
\mbox{}{\tau_8}}{{P_1(\ta)}}}\,,
\end{equation}
\begin{equation}
\label{VmuF4rat}
V_{\mu}\ =\ -\ \frac{1}{8}\
            \frac{\left({\tau_2}^{4} - 48\,{\tau_2}\,{\tau_6} + 192\,{\tau_8}
\right)
            \bigg( {\tau_2}^{3}{\tau_8} - 3\,{\tau_2} ({\tau_6}^{2} + 8\,{\tau_{12}}) + 48\,{\tau_6}\,{\tau_8} \bigg)}{{P_2(\ta)}} \ .
\end{equation}

Thus, the original potential (\ref{HamF4rat}) is a rational (meromorphic) function in $\ta$-variables. From the expressions (\ref{VnuF4rat}) and (\ref{VmuF4rat}), it is clear that the singularities of the potential are defined by the zeroes of the (ground state) factors $P_1(\ta), P_2(\ta)$ (cf.(\ref{DD})), and therefore they define the boundaries of the configuration space.

\subsection{$sl(2)$ quasi-exactly-solvable $F_4$-invariant rational model}

It can be shown \cite{BLT-rational} that the polynomial generalization of the potential $V_{F_4}^{\rm (rat)}$ (\ref{VF4rat}),
\[
V_{F_4}^{\rm (qes)}(\ta)\ =\ V_{F_4}^{\rm (rat)}+a^2 \ta_2^3 + 2a\om \ta_2^2 + 2a[2k-\ga+3(4\mu+4\nu+1)]\ta_2 + \frac{2\ga[\ga-12\mu-12\nu-1)]}{\ta_2}\ ,
\]
where $a, \ga$ are real parameters, $k$ is non-negative integer, $\mu, \nu$ are given in (\ref{gg}), leads to the so-called $sl(2)$ quasi-exactly-solvable $F_4$-invariant rational Hamiltonian
\[
   H^{\rm (qes)}_{F_4} \ =\ -\De_g (\ta) + V_{F_4}^{\rm (qes)}(\ta)\ .
\]
This Hamiltonian has a single finite-dimensional invariant subspace
${\mathcal P}_{k}^{({1},{2},{2},{3})}$ (see (\ref{minflagF4rat})). Hence,
$(k+1)$ eigenfunctions have the form
\begin{equation}
\label{Psiqes}
 \Psi_{k}^{{\rm (qes)}}\ =\ \left(\De_- \De_+\right)^{\nu}
 \left(\De_0 \De \right)^{\mu} \ta_2^{\ga}
 P_k (\ta_2)\ \exp\left[{-\om\ta_2 - \frac{a}{4} \ta_2^2}\right]
\end{equation}
\[
 \ =\ \ta_2^{\ga} P_k (\ta_2)\ \exp{\bigg( - \frac{a}{4} \ta_2^2\bigg)} \Psi_{0}
 \ \equiv \ P_k (\ta_2) \Psi_{0}^{({\rm qes})}
 \ ,
\]
where $P_k (\ta_2)$ is a $k$th degree polynomial, and can be found algebraically.
It can be easily checked that
\[
   \mathfrak{h}_{F_4}^{\rm (qes)}(\ta)\ =\ {(\Psi_0^{({\rm qes})})}^{-1} \, { \left[ H_{F_4}^{({\rm qes})} -E_0 \right] }\,\,{\Psi_0^{({\rm qes})}}\ ,
\]
is algebraic operator.

\subsection{Particular integral}

Let us consider a certain Euler-Cartan operator in $\ta$-variables

\[
    {\mathcal J}^0\ =\ \ta_1\frac{\pa}{\pa \ta_1} + 2\ta_2\frac{\pa}{\pa \ta_2}+
    2\ta_3\frac{\pa}{\pa \ta_3} + 3\ta_4\frac{\pa}{\pa \ta_4}\ .
\]
and form
\begin{equation}
\label{integral}
       i_{par}^{(k)}(\ta)\ =\ \prod_{j=0}^k ({\mathcal J}^0 - j)\ .
\end{equation}
It is easy to check that
\[
  i_{par}^{(k)}(\ta)\ :\ {\mathcal P}_{k}^{({1},{2},{2},{3})} \ \mapsto \ 0\ .
\]
So, ${\mathcal P}_{k}^{({1},{2},{2},{3})}$ is the space of zero modes for $i_{par}^{(k)}(\ta)$.
Hence, this operator commutes with both $\mathfrak{h}_{F_4}^{\rm rat}$ and $\mathfrak{h}_{F_4}^{\rm qes}$,
\[
  [\mathfrak{h}_{F_4}^{\rm rat}\ (\mathfrak{h}_{F_4}^{\rm qes})\ ,\ i_{par}^{(k)}(\ta)]:\ {\mathcal P}_{k}^{({1},{2},{2},{3})} \ \mapsto \ 0\ .
\]
The operator $i_{par}^{(k)}$ is called a particular integral.


\section{$F_4$ Trigonometric Model}

The Hamiltonian of the quantum $F_4$ trigonometric model is given by
\footnote{In previous studies \cite{blt, BLT-trigonometric} the analysis of the $F_4$-trigonometric Hamiltonian was done in the dual root space.}
\begin{eqnarray}
 H_{F_4}^{\rm trig} &=& -\frac{1}{2} {\sum_{k=1}^4 }\frac{\pa^{2}}{\pa x _{k }^{2}}\ +\
  g _{l }\frac{\beta^{2}}{4}
 {\sum_{1\le j <i \le 4} }\left(\frac{1}{\sin^{2}\frac{\beta}{2}\left(x_{i }-x_{j }\right)}\ +\ \frac{1}{\sin^{2}\frac{\beta }{2}\left(x_{i}+x_{j}\right)}\right)
 \label{Hexpanded} \\
 & & +\ g_{s}\frac{\beta^{2}}{8}{\sum_{i =1}^4}\frac{1}{\sin ^{2}\frac{\beta }{2} x_{i}}
  \ +\ g_{s} \frac{\beta^{2}}{8} \sum_{\nu_j =0,1}
  \frac{1}{\sin^{2}\frac{\beta}{4}\bigg(x_{1}+\left(-\right)^{\nu _{2}}x_{2}\ +\ \left(-\right)^{\nu_{3}}x_{3}+\left(-\right)^{\nu_{4}}x_{4}\bigg)}
   \non\,,
\end{eqnarray}
where  $g_s, g_{\ell}$  are  coupling constants assigned to the potential terms associated with short and long roots respectively. The configuration space is the $F_4$-Weyl Alcove.
The Hamiltonian (\ref{Hexpanded}) is invariant with respect to $F_4$-Weyl group transformations. The parameter $\beta$ (the inverse of the period) is a parameter introduced for convenience in such a way that
\[
 \lim_{\beta\to 0} H_{F_4}^{\rm trig} \to H_{F_4}^{\rm rat} (\om=0)\,,
\]
{\it i.e.} the rational model (without the harmonic oscillator term) is reproduced. If $\beta \rar i \beta$, the Hamiltonian (\ref{Hexpanded}) becomes the Hamiltonian of the $F_4$-invariant hyperbolic model.

The ground state eigenfunction of the Hamiltonian  (\ref{Hexpanded}) is given by
\begin{equation}
\label{GSF4trig}
 \psi_{0}(x) =  (\De_{+}\,\De_{-})^\nu \,(\De\,\De_{0})^\mu \ ,
\end{equation}
where
\begin{equation}
 \label{DeltapmTrig}
\De_{+}\,\De_{-} = \prod_{\al_{\rm long}\in {\mathcal F_4}^{+}} \sin \frac{\beta}{2} (\al_{\rm long}\cdot x)\,,
\end{equation}
 \begin{equation}
 \label{Delta0Trig}
\De\,\De_{0} = \prod_{\al_{\rm short}\in {\mathcal F_4}^{+}} \sin \frac{\beta}{2} (\al_{\rm short}\cdot x)\,.
 \end{equation}
and the exponents $\mu,\nu$ in the ground state eigenfunction (\ref{GSF4trig}) are connected to the coupling constants $g_s, g_{\ell}$ by the relations
\[
   g_{\ell}=\nu(\nu-1) \quad , \quad g_s=\mu(\mu-1) \ .
\]
The ground state eigenvalue is then
\[
E_{0}=  \frac{1}{2}\,{\beta}^{2} \left(14\,{\nu}^{2}  +18\,\nu\,\mu + 7\,{\mu}^{2}\right)\,,
\]
which can be written in terms of the {\it deformed Weyl vector\footnote{The standard
definition for the Weyl vector is $\rho= \frac{1}{2}\sum_{\alpha\in {\mathcal R}^{+}} \alpha$.}}
\begin{equation}
\label{deformedweyl}
 \varrho = \frac{1}{2} \left(
 \mu \sum_{\al_{\rm short} \in {\mathcal F}_4^{+}} \al_{\rm short} +
 \nu \sum_{\al_{\rm long} \in {\mathcal F}_4^{+}} \al_{\rm long}  \right)\,,
\end{equation}
as
\[
 E_0 = \frac{1}{2}\,{\beta}^{2} \varrho^2\,.
\]

All eigenfunctions $\psi_n(x)$ of the Hamiltonian operator  (\ref{Hexpanded}) have a factorized form
\[
 \psi_n(x) = \psi_0(x)\, \phi_n(x) \,.
\]
The  functions $\phi_n(x)$  are eigenfunctions of the gauge rotated Hamiltonian operator defined by
\begin{equation}
\label{hF4trig}
 \mathfrak{h}_{F_4}^{\rm trig} = - \frac{2}{\beta^2} \psi_0^{-1}\,\,\Big[ H_{F_4}^{\rm trig} - E_0\Big]\, \psi_0 \,.
\end{equation}
Then, the eigenvalue problem becomes
\[
  \mathfrak{h}_{F_4}^{\rm trig}\,  \phi_n = \epsilon_n \phi_n\,,
\]
where $\ep_n =  -\frac{2}{\beta^2} (E_n-E_0)$.

Let us introduce the Weyl-invariant trigonometric coordinates
(see \cite{BLT-trigonometric})
\begin{equation}
\label{F4trigtauvars}
\tau_a \equiv \sum_{\om \in \Om_a} e^{i\,\beta(\om\cdot x)}\,, \qquad {a=1,2,3,4}
\end{equation}
where $ \Om_a$ is the Weyl orbit generated by the fundamental
weight $\varpi_a$:
\footnote{Bourbaki numbering: \( \varpi_1 =  e_1 + e_2 ,\quad
\varpi_2 =  2e_1 + e_2 + e_3 , \quad
\varpi_3 = \frac{1}{2} (3e_1 + e_2 + e_3 + e_4),\quad
\varpi_4 = e_1\), \hbox{(see \cite{Bourbaki:2002})}.}
\begin{equation}
\label{fundweights}
\varpi_1 =  e_4 ,\quad
\varpi_2 =  e_3 + e_4 , \quad
\varpi_3 = \frac{1}{2} (e_1 + e_2 + e_3 + 3e_4),\quad
\varpi_4 = e_2 + e_3 + 2e_4.
\end{equation}
The corresponding orbit sizes are $|\Om_a| = 24,24, 96, 96$ for $a=1,2,3,4$ respectively.
The explicit form of variables $\tau_a, a=1,2,3,4$ is given at Appendix C.

Changing variables $\{x\} \to \{\tau\}$ we obtain the gauge-rotated Hamiltonian (\ref{hF4trig})
\begin{equation}
\label{hF4algebraic}
  \mathfrak{h}_{F_4}^{\rm trig}  = \sum_{a,b=1}^{4} \mathcal{A}_{ab} \frac{\partial^2}{\partial \tau_a\, \partial \tau_b} +
                                   \sum_{a=1}^{4} \mathcal{B}_{a} \frac{\partial}{\partial \tau_a}\,,
\end{equation}
with coefficient functions
\begin{eqnarray*}
\mathcal{A}_{{1,1}}\,  &=&  \,    -{\tau_{{1}}}^{2}+12\,\tau_{{1}}+6\,\tau_{{2}}+\tau_{{3}}
               +48\ ,
   \\
\mathcal{A}_{{1,2}}\,  &=&  \,    -\tau_{{1}}\tau_{{2}}+12\,\tau_{{1}}+3\,\tau_{{3}}\ ,
   \\
\mathcal{A}_{{1,3}}\,  &=&  \,    12\,{\tau_{{1}}}^{2}+4\,\tau_{{1}}\tau_{{2}}-3/2\,\tau_{{1}}\tau_{{3}}-96\,\tau_{{1}}
      -42\,\tau_{{2}}-24\,\tau_{{3}} +3/2\,\tau_{{4}}-288\ ,
   \\
\mathcal{A}_{{1,4}}\,  &=&  \,    4\,\tau_{{1}}\tau_{{2}}-2\,\tau_{{1}}\tau_{4}+2\,\tau_{{2}}\tau_{3}
      -48\,\tau_{1}-12\,\tau_{3}\ ,
   \\
\mathcal{A}_{{2,2}}\,  &=&  \,    12\,{\tau_{{1}}}^{2}-2\,{\tau_{{2}}}^{2}-96\,\tau_{{1}}-48\,\tau_{{2}}-24\,\tau_{{3}}
      +2\,\tau_{4}-192 \ ,
   \\
\mathcal{A}_{{2,3}}\,  &=&  \,    -24\,{\tau_{{1}}}^{2}-4\,\tau_{{1}}\tau_{{2}}+3\,\tau_{{1}}\tau_{{3}}-2\,\tau_{{2}}\tau_{{3}}
          +240\,\tau_{1} + 108\,\tau_{{2}}+60\,\tau_{{3}}-9\,\tau_{{4}}+576\ ,
   \\
\mathcal{A}_{2,4}\,  &=&  \,    -24\,{\tau_{{1}}}^{3}-4\,{\tau_{{1}}}^{2}\tau_{{2}}+96\,{\tau_{{1}}}^{2}+104\,\tau_{{1}}\tau_{{2}}
                            +72\,\tau_{{1}}\tau_{{3}} -6\,\tau_{{1}}\tau_{{4}}+12\,{\tau_{{2}}}^{2}+8\,\tau_{{2}}\tau_{{3}}
                       \\&& \hspace{20pt}  -3\,\tau_{{2}}\tau_{{4}}+3\,{\tau_{{3}}}^{2}+1536\,\tau_{{1}} +480\,\tau_{{2}}+288\,\tau_{{3}}-48\,\tau_{{4}}+2304\ ,
   \\
\mathcal{A}_{{3,3}}\,  &=&  \,    12\,\tau_{1}^{3}+4\,{\tau_{1}}^{2}\tau_{2}-96\,{\tau_{1}}^{2}-60\,\tau_{1}\tau_{2}-
      36\,\tau_{1}\tau_{3}
      +\tau_{1}\tau_{4}-4\,\tau_{{2}}\tau_{3}-3\,{\tau_{3}}^{2}
      \\&& \hspace{20pt} -384\,\tau_{1}-48\,\tau_{2}-48\,\tau_{3}+12\,\tau_{4}\ ,
   \\
\mathcal{A}_{{3,4}}\,  &=&  \,   -16\,{\tau_{{1}}}^{2}\tau_{{2}}+2\,\tau_{{1}}\tau_{{2}}\tau_{{3}}+96\,{\tau_{{1}}}^{2}
     +144\,\tau_{{1}}\tau_{{2}} -12\,\tau_{{1}}\tau_{{3}}-8\,\tau_{{1}}\tau_{{4}}+72\,{\tau_{{2}}}^{2}
     \\ & & \hspace{20pt}
     +32\,\tau_{{2}}\tau_{{3}}-6\,\tau_{{2}}\tau_{{4}}
     -4\,\tau_{3}\tau_{4}-960\,\tau_{1}-48\,\tau_{2}-240\,\tau_{3}+36\,\tau_{4}-2304\ ,
    \\
\mathcal{A}_{{4,4}}\,  &=&  \, 9216+2880\,\tau_{1}\tau_{{2}}+576\,\tau_{1}\tau_{3}+512\,\tau_{2}\tau_{3}
    +16\,\tau_{1}\tau_{4}-24\,\tau_{2}\tau_{4}-96\,{\tau_{1}}^{2}\tau_{2}
    \\ & & \hspace{10pt}
    +16\, \tau_{3} \tau_{4} + 2\,\tau_{2}{\tau_{3}}^{2}-8\,{\tau_{1}}^{2}\tau_{4}
    +48\,\tau_{{1}}{\tau_{{2}}}^{2}
    -16\,{\tau_{{1}}}^{3}\tau_{{2}} +7680\,\tau_{{1}}+6144\,\tau_{{2}}+1152\,\tau_{{3}}
    \\ & & \hspace{20pt}
   +96\,\tau_{{4}}+1344\,{\tau_{1}}^{2}+864\,{\tau_{2}}^{2}-192\,{\tau_{1}}^{3}
   +24\,{\tau_{3}}^{2}-6\,{\tau_{4}}^{2}+48\,\tau_{1}\tau_{2}\tau_{3}
   -4\,\tau_{1}\tau_{2}\tau_{4} \ ,
\end{eqnarray*}
and
\begin{eqnarray*}
\mathcal{B}_{1}\, &=& \,  - \tau_{1}    - 24\,\mu  - \left( 5\,\mu
         + 6\,\nu \right) \tau_{1}\ ,
\\
\mathcal{B}_{2}\, &=& \, -2\, \tau_{2}   - 48\,\nu  - 6\,\mu\,\tau_{{1}}
        - \left( 6\,\mu  + 10\,\nu \right) \tau_{2}\ ,
\\
\mathcal{B}_{3}\, &=& \, -3\, \tau_{3}  -24 \left(\,\mu + \,\nu \right) \tau_{1}
- 12\,\mu\,\tau_{2} -3\, \left( \,3 \mu + \,4\nu \right) \tau_{3}  \ ,
\\
\mathcal{B}_{4}\, &=& \, -6\, \tau_{4} +  576\,\nu + 24\,\left( \mu
     + \,8\nu \right) \tau_{1} - 24\,\left( \,\mu - 4\,\nu \right) \tau_{2}
     + 48\,\nu\,\tau_{3} - 6\,\left( \,2\mu + \,3\nu \right) \tau_{4}
 \\&& \hspace{20pt}
-24\,\nu\,{\tau_{1}}^{2}-4\,\mu\,\tau_{{1}}\tau_{2}\ .
\end{eqnarray*}

\subsection{Flat Metric of the $F_4$-trigonometric Model.}

It is evident that gauge rotated Hamiltonian $\mathfrak{h}_{F_4}^{\rm trig}$ (\ref{hF4trig}) in algebraic form (\ref{hF4algebraic}) can be written as
\begin{equation}
\label{hF4trigD}
 \beta^2\ \mathfrak{h}_{F_4}^{\rm trig}\ =\ \De_g\ +\ {\mathcal C}^b \frac{\pa}{\pa \ta_{b}}\ ,
\end{equation}
where the Laplace-Beltrami operator
\begin{equation}
\label{LBtrig}
    \De_g\ =\ \frac{1}{g^{1/2}}\frac{\pa}{\pa \ta_{a}}\ (g^{a b} g^{1/2})\ \frac{\pa}{\pa \ta_{b}}\ =\ g^{a b} \frac{\pa^2}{\pa \ta_{a}\pa \ta_{b}}\ +\ g^{b} \frac{\pa}{\pa \ta_{b}}\ ,\ g^{b} \equiv \frac{1}{g^{1/2}}\frac{\pa}{\pa \ta_{a}}\ (g^{a b} g^{1/2}) \ ,
\end{equation}
with $a,b=1,2,3,4$ and $g$ is determinant. The symmetric (in $a \lrar b$) metric, $g^{a b}=g^{b a}$ has polynomial in $\ta$ matrix elements

\begin{equation}
g^{a b}\ = \left| \begin{array}{cccc}
 \mathcal{A}_{1,1} & \mathcal{A}_{1,2} & \mathcal{A}_{1,3} & \mathcal{A}_{1,4}\\
     &    \\
  & \mathcal{A}_{2,2} & \mathcal{A}_{2,3} & \mathcal{A}_{2,4}\\
     &    \\
  &  & \mathcal{A}_{3,3} & \mathcal{A}_{3,4}\\
     &    \\
   &  &
   &  \mathcal{A}_{4,4}
 \end{array} \right|\ .
\end{equation}
This flat metric has a special property: any component $g^b$ of the vector $\vec{g}$ in (\ref{LBtrig}) is also polynomial in $\ta$
\[
    - \vec{g}\ =\ (\ta_1,\ 2\ta_2,\ 3\ta_3,\ 6\ta_4)\ .
\]
The same property has the vector ${\mathcal C}^b$ (see (\ref{hF4trigD})) which characterizes the interaction in (\ref{Hexpanded})
\[
     -\vec{\mathcal C}\ =\ \bigg(
     24\mu  +(5\mu  + 6\nu) \ta_{1} ,\
     48\nu  + 6\mu\ta_{1} + 2(3\mu + 5\nu)\ta_2 ,\
     24(\mu + \nu)\ta_1 + 12\mu\ta_{2}+3(3\mu + 4\nu)\ta_3 ,\
\]
\[
     -576\nu - 24(\mu +8\nu)\ta_{1} + 24(\mu - 4\nu)\ta_{2}
     -48\nu\ta_3 + 6(2\mu + 3\nu)\ta_4
     +24\nu{\ta_1}^2+4\mu\ta_1\tau_2
        \bigg)\ .
\]

\subsection{Ground state eigenfunction and potential of $F_4$-trigonometric model in $\tau$-variables}

The ground state eigenfunction (\ref{GSF4trig}) and the original potential in (\ref{Hexpanded}) can be written in terms of the invariant variables (\ref{F4trigtauvars}). After straightforward but very cumbersome calculations we find that
\[
(\Delta_{+}\Delta_{-})^2 = \left(\frac{1}{2}\right)^{24} P_{1}(\tau)\,, \qquad
(\Delta\Delta_0)^2 = \left(\frac{1}{2}\right)^{24} P_{2}(\tau)\,,
\]
where
\begin{eqnarray*}
P_1(\tau) &=&-1728\,{\tau_{{1}}}^{6}-1728\,{\tau_{{1}}}^{5}\tau_{{2}}
-432\,{\tau_{{1}}}^{4}{\tau_{{2}}}^{2}+32\,{\tau_{{1}}}^{3}{\tau_{{2}}}^{3}
+16\,{\tau_{{1}}}^{2}{\tau_{{2}}}^{4}+20736\,{\tau_{{1}}}^{5}
\\&&
+34560\,{\tau_{{1}}}^{4}\tau_{{2}}+10368\,{\tau_{{1}}}^{4}\tau_{{3}}
-864\,{\tau_{{1}}}^{4}\tau_{{4}}+18432\,{\tau_{{1}}}^{3}{\tau_{{2}}}^{2}
+8640\,{\tau_{{1}}}^{3}\tau_{{2}}\tau_{{3}}
\\&&
-576\,{\tau_{{1}}}^{3}\tau_{{2}}\tau_{{4}}
+432\,{\tau_{{1}}}^{3}{\tau_{{3}}}^{2}+2976\,{\tau_{{1}}}^{2}{\tau_{{2}}}^{3}
+1728\,{\tau_{{1}}}^{2}{\tau_{{2}}}^{2}\tau_{{3}}
-72\,{\tau_{{1}}}^{2}{\tau_{{2}}}^{2}\tau_{{4}}
\\&&
+216\,{\tau_{{1}}}^{2}\tau_{{2}}{\tau_{{3}}}^{2}
-224\,\tau_{{1}}{\tau_{{2}}}^{4}-96\,\tau_{{1}}{\tau_{{2}}}^{3}\tau_{{3}}
+8\,\tau_{{1}}{\tau_{{2}}}^{3}\tau_{{4}}-64\,{\tau_{{2}}}^{5}
-32\,{\tau_{{2}}}^{4}\tau_{{3}}
\\&&
-4\,{\tau_{{2}}}^{3}{\tau_{{3}}}^{2}+103680\,{\tau_{{1}}}^{4}
+6912\,{\tau_{{1}}}^{3}\tau_{{2}}-34560\,{\tau_{{1}}}^{3}\tau_{{3}}
+1728\,{\tau_{{1}}}^{3}\tau_{{4}}
\\&&
-88128\,{\tau_{{1}}}^{2}{\tau_{{2}}}^{2}-79488\,{\tau_{{1}}}^{2}\tau_{{2}}\tau_{{3}}
+5184\,{\tau_{{1}}}^{2}\tau_{{2}}\tau_{{4}}-18144\,{\tau_{{1}}}^{2}{\tau_{{3}}}^{2}
+2592\,{\tau_{{1}}}^{2}\tau_{{3}}\tau_{{4}}
\\&&
-108\,{\tau_{{1}}}^{2}{\tau_{{4}}}^{2}
-45888\,\tau_{{1}}{\tau_{{2}}}^{3}-43200\,\tau_{{1}}{\tau_{{2}}}^{2}\tau_{{3}}
+2592\,\tau_{{1}}{\tau_{{2}}}^{2}\tau_{{4}}-13392\,\tau_{{1}}\tau_{{2}}{\tau_{{3}}}^{2}
\\&&
+1296\,\tau_{{1}}\tau_{{2}}\tau_{{3}}\tau_{{4}}
-36\,\tau_{{1}}\tau_{{2}}{\tau_{{4}}}^{2}-1296\,\tau_{{1}}{\tau_{{3}}}^{3}
+108\,\tau_{{1}}{\tau_{{3}}}^{2}\tau_{{4}}-6384\,{\tau_{{2}}}^{4}
-6592\,{\tau_{{2}}}^{3}\tau_{{3}}
\\&&
+328\,{\tau_{{2}}}^{3}\tau_{{4}}
-2520\,{\tau_{{2}}}^{2}{\tau_{{3}}}^{2}+144\,{\tau_{{2}}}^{2}\tau_{{3}}\tau_{{4}}
+{\tau_{{2}}}^{2}{\tau_{{4}}}^{2}-432\,\tau_{{2}}{\tau_{{3}}}^{3}
+18\,\tau_{{2}}{\tau_{{3}}}^{2}\tau_{{4}}
\\&&
-27\,{\tau_{{3}}}^{4}
-774144\,{\tau_{{1}}}^{3}-1465344\,{\tau_{{1}}}^{2}\tau_{{2}}
-663552\,{\tau_{{1}}}^{2}\tau_{{3}}+62208\,{\tau_{{1}}}^{2}\tau_{{4}}
\\&&
-787968\,\tau_{{1}}{\tau_{{2}}}^{2}-566784\,\tau_{{1}}\tau_{{2}}\tau_{{3}}
+48384\,\tau_{{1}}\tau_{{2}}\tau_{{4}}
-103680\,\tau_{{1}}{\tau_{{3}}}^{2}+17280\,\tau_{{1}}\tau_{{3}}\tau_{{4}}
\\&&
-864\,\tau_{{1}}{\tau_{{4}}}^{2}-129024\,{\tau_{{2}}}^{3}
-119808\,{\tau_{{2}}}^{2}\tau_{{3}}+9024\,{\tau_{{2}}}^{2}\tau_{{4}}
-36288\,\tau_{{2}}{\tau_{{3}}}^{2}+4608\,\tau_{{2}}\tau_{{3}}\tau_{{4}}
\\&&
-192\,\tau_{{2}}{\tau_{{4}}}^{2}-3456\,{\tau_{{3}}}^{3}
+432\,{\tau_{{3}}}^{2}\tau_{{4}}-4\,{\tau_{{4}}}^{3}-5308416\,{\tau_{{1}}}^{2}
-4866048\,\tau_{{1}}\tau_{{2}}
\\&&
-1990656\,\tau_{{1}}\tau_{{3}}
+221184\,\tau_{{1}}\tau_{{4}}-1096704\,{\tau_{{2}}}^{2}
-774144\,\tau_{{2}}\tau_{{3}}+78336\,\tau_{{2}}\tau_{{4}}
\\&&
-138240\,{\tau_{{3}}}^{2}+27648\,\tau_{{3}}\tau_{{4}}-
1728\,{\tau_{{4}}}^{2}-10616832\,\tau_{{1}}-4423680\,\tau_{{2}}
-1769472\,\tau_{{3}}
\\&&
+221184\,\tau_{{4}}-7077888\ ,
\end{eqnarray*}
\begin{eqnarray*}
P_2(\tau) &=&
-16\,{\tau_{1}}^{5}+48\,{\tau_{1}}^{3}\tau_{2}
+112\,{\tau_{{1}}}^{3}\tau_{{3}}-4\,{\tau_{{1}}}^{3}\tau_{{4}}
+{\tau_{{1}}}^{2}{\tau_{{3}}}^{2}+4608\,{\tau_{{1}}}^{3}
+1728\,{\tau_{{1}}}^{2}\tau_{{2}}
\\&&
+384\,{\tau_{{1}}}^{2}\tau_{{3}}-144\,{\tau_{{1}}}^{2}\tau_{{4}}
-216\,\tau_{{1}}\tau_{{2}}\tau_{{3}}-192\,\tau_{{1}}{\tau_{{3}}}^{2}
+18\,\tau_{{1}}\tau_{{3}}\tau_{{4}}-4\,{\tau_{{3}}}^{3}-18432\,{\tau_{{1}}}^{2}
\\&&
-20736\,\tau_{{1}}\tau_{{2}}-14976\,\tau_{{1}}\tau_{{3}}
+1728\,\tau_{{1}}\tau_{{4}}-3888\,{\tau_{{2}}}^{2}-5184\,\tau_{{2}}\tau_{{3}}
+648\,\tau_{{2}}\tau_{{4}}-1728\,{\tau_{{3}}}^{2}
\\&&
+432\,\tau_{3}\tau_{4}-27\,{\tau_{4}}^{2}-110592\,\tau_{1}
-41472\,\tau_{{2}}-27648\,\tau_{{3}}+3456\,\tau_{4}-110592\ .
\end{eqnarray*}
Thus, the  unnormalized ground state eigenfunction (\ref{GSF4trig}) becomes
\begin{equation}
\label{Psi0intauF4trig}
\psi_0(\ta)\ =\ {P_1}(\ta)^{\nu/2} \,{P_2}(\ta)^{\mu/2}\ ,
\end{equation}
(cf. (\ref{GSF4trig})).

It is evident that the original potential of $F_4$-invariant trigonometric model (\ref{Hexpanded}) can be written in terms of  $\ta$ variables. The simplest way to do it is to invert the gauge rotation (\ref{hF4trig}) with the ground state eigenfunction (\ref{Psi0intauF4trig}) of the algebraic operator (\ref{hF4algebraic}). After quite sophisticated technically, tedious calculations we get
\begin{equation}
\label{Vtau}
V_{F_4}^{\rm trig}(\tau) =  - {g_{\ell}\beta^2} \frac{\mathcal{N}_1 (\tau)}{P_1(\tau)}    -\frac{g_{s}\beta^2}{2} \frac{\mathcal{N}_2(\tau)}{P_2(\tau)}  \ ,
\end{equation}
 where
\begin{eqnarray*}
\mathcal{N}_1 &=& \left(1728\,{\tau_{{1}}}^{6}+1728\,{\tau_{{1}}}^{5}\tau_{{2}}
+432\,{\tau_{{1}}}^{4}{\tau_{{2}}}^{2}-8\,{\tau_{{1}}}^{3}{\tau_{{2}}}^{3}
-8\,{\tau_{{1}}}^{2}{\tau_{{2}}}^{4}-20736\,{\tau_{{1}}}^{5}
\right.\\&&\left.
-34560\,{\tau_{{1}}}^{4}\tau_{{2}}-10368\,{\tau_{{1}}}^{4}\tau_{{3}}
+864\,{\tau_{{1}}}^{4}\tau_{{4}}-19296\,{\tau_{{1}}}^{3}{\tau_{{2}}}^{2}
-8640\,{\tau_{{1}}}^{3}\tau_{{2}}\tau_{{3}}
\right.\\&&\left.
+504\,{\tau_{{1}}}^{3}\tau_{{2}}\tau_{{4}}
-432\,{\tau_{{1}}}^{3}{\tau_{{3}}}^{2}-3456\,{\tau_{{1}}}^{2}{\tau_{{2}}}^{3}
-1728\,{\tau_{{1}}}^{2}{\tau_{{2}}}^{2}\tau_{{3}}
+60\,{\tau_{{1}}}^{2}{\tau_{{2}}}^{2}\tau_{{4}}
\right.\\&&\left.
-216\,{\tau_{{1}}}^{2}\tau_{{2}}{\tau_{{3}}}^{2}+88\,\tau_{{1}}{\tau_{{2}}}^{4}
+24\,\tau_{{1}}{\tau_{{2}}}^{3}\tau_{{3}}-2\,\tau_{{1}}{\tau_{{2}}}^{3}\tau_{{4}}
+48\,{\tau_{{2}}}^{5}+16\,{\tau_{{2}}}^{4}\tau_{{3}}
\right.\\&&\left.
+{\tau_{{2}}}^{3}{\tau_{{3}}}^{2}-103680\,{\tau_{{1}}}^{4}
+34560\,{\tau_{{1}}}^{3}\tau_{{3}}+96192\,{\tau_{{1}}}^{2}{\tau_{{2}}}^{2}
+79488\,{\tau_{{1}}}^{2}\tau_{{2}}\tau_{{3}}
\right.\\&&\left.
-4176\,{\tau_{{1}}}^{2}\tau_{{2}}\tau_{{4}}
+18144\,{\tau_{{1}}}^{2}{\tau_{{3}}}^{2}
-2592\,{\tau_{{1}}}^{2}\tau_{{3}}\tau_{{4}}
+72\,{\tau_{{1}}}^{2}{\tau_{{4}}}^{2}+50016\,\tau_{{1}}{\tau_{{2}}}^{3}
\right.\\&&\left.
+45792\,\tau_{{1}}{\tau_{{2}}}^{2}\tau_{{3}}
-2496\,\tau_{{1}}{\tau_{{2}}}^{2}\tau_{{4}}
+13392\,\tau_{{1}}\tau_{{2}}{\tau_{{3}}}^{2}
-1080\,\tau_{{1}}\tau_{{2}}\tau_{{3}}\tau_{{4}}
+18\,\tau_{{1}}\tau_{{2}}{\tau_{{4}}}^{2}
\right.\\&&\left.
+1296\,\tau_{{1}}{\tau_{{3}}}^{3}
-108\,\tau_{{1}}{\tau_{{3}}}^{2}\tau_{{4}}+6912\,{\tau_{{2}}}^{4}
+7072\,{\tau_{{2}}}^{3}\tau_{{3}}-352\,{\tau_{{2}}}^{3}\tau_{{4}}
+2628\,{\tau_{{2}}}^{2}{\tau_{{3}}}^{2}
\right.\\&&\left.
-120\,{\tau_{{2}}}^{2}\tau_{{3}}\tau_{{4}}+432\,\tau_{{2}}{\tau_{{3}}}^{3}
-9\,\tau_{{2}}{\tau_{{3}}}^{2}\tau_{{4}}+27\,{\tau_{{3}}}^{4}
+774144\,{\tau_{{1}}}^{3}+1423872\,{\tau_{{1}}}^{2}\tau_{{2}}
\right.\\&&\left.
+663552\,{\tau_{{1}}}^{2}\tau_{{3}}-72576\,{\tau_{{1}}}^{2}\tau_{{4}}
+785664\,\tau_{{1}}{\tau_{{2}}}^{2}+546048\,\tau_{{1}}\tau_{{2}}\tau_{{3}}
-52992\,\tau_{{1}}\tau_{{2}}\tau_{{4}}
\right.\\&&\left.
+103680\,\tau_{{1}}{\tau_{{3}}}^{2}
-22464\,\tau_{{1}}\tau_{{3}}\tau_{{4}}+1584\,\tau_{{1}}{\tau_{{4}}}^{2}
+134208\,{\tau_{{2}}}^{3}+120960\,{\tau_{{2}}}^{2}\tau_{{3}}
\right.\\&&\left.
-9936\,{\tau_{{2}}}^{2}\tau_{{4}}+35424\,\tau_{{2}}{\tau_{{3}}}^{2}
-5184\,\tau_{{2}}\tau_{{3}}\tau_{{4}}+396\,\tau_{{2}}{\tau_{{4}}}^{2}
+3456\,{\tau_{{3}}}^{3}-648\,{\tau_{{3}}}^{2}\tau_{{4}}
\right.\\&&\left.
+72\,\tau_{{3}}{\tau_{{4}}}^{2}+{\tau_{{4}}}^{3}
+5308416\,{\tau_{{1}}}^{2}+4534272\,\tau_{{1}}\tau_{{2}}
+1990656\,\tau_{{1}}\tau_{{3}}-304128\,\tau_{{1}}\tau_{{4}}
\right.\\&&\left.
+1022976\,{\tau_{{2}}}^{2}+718848\,\tau_{{2}}\tau_{{3}}
-96768\,\tau_{{2}}\tau_{{4}}+138240\,{\tau_{{3}}}^{2}-41472\,\tau_{{3}}\tau_{{4}}
+4032\,{\tau_{{4}}}^{2}
\right.\\&&\left.
+10616832\,\tau_{{1}}+3981312\,\tau_{{2}}
+1769472\,\tau_{{3}}-331776\,\tau_{{4}}+7077888 \right)\ ,
\end{eqnarray*}
and
\begin{eqnarray*}
\mathcal{N}_2 &=&  \left(
 12\,{\tau_{{1}}}^{5}-4\,{\tau_{{1}}}^{4}\tau_{{2}}
+288\,{\tau_{{1}}}^{4}+84\,{\tau_{{1}}}^{3}\tau_{{2}}
-100\,{\tau_{{1}}}^{3}\tau_{{3}}+{\tau_{{1}}}^{3}\tau_{{4}}
+24\,{\tau_{{1}}}^{2}\tau_{{2}}\tau_{{3}}-9216\,{\tau_{{1}}}^{3}\right.
\\&&
-3600\,{\tau_{{1}}}^{2}\tau_{{2}}-1584\,{\tau_{{1}}}^{2}\tau_{{3}}
+252\,{\tau_{{1}}}^{2}\tau_{{4}}-180\,\tau_{{1}}\tau_{{2}}\tau_{{3}}
+180\,\tau_{{1}}{\tau_{{3}}}^{2}-9\,\tau_{{1}}\tau_{{3}}\tau_{{4}}
\\&&
-36\,\tau_{2}{\tau_{3}}^{2}+{\tau_{3}}^{3}
+34560\,{\tau_{1}}^{2}+31104\,\tau_{1}\tau_{2}
+25920\,\tau_{1}\tau_{3}-2592\,\tau_{1}\tau_{4}
+3888\,{\tau_{2}}^{2}
\\&&
+7776\,\tau_{2}\tau_{3}
-648\,\tau_{{2}}\tau_{{4}}+3168\,{\tau_{{3}}}^{2}-648\,\tau_{{3}}\tau_{{4}}
+27\,{\tau_{4}}^{2}+165888\,\tau_{1}+41472\,\tau_{2}
\\&&
\left.
+41472\,\tau_{3}-3456\,\tau_{4}+110592 \right)\ .
\end{eqnarray*}
Then, the $F_4$-trigonometric potential written in $\ta$-variables is a rational (meromorphic) function! From the structure of the potential given in  (\ref{Vtau}) it is clear that the singularities of the potential are defined by the zeroes of the (ground state) factors $P_1(\ta), P_2(\ta)$, and therefore they lie on the boundaries of the configuration space.

\subsection{Exact Solvability (Invariant subspaces)}

The Hamiltonian (\ref{hF4algebraic}) has an infinite number of invariant polynomial subspaces
\[
 \mathcal{P}^{(p_1,p_2,p_3,p_4)}_n \equiv
 \left\{
 \tau_1^{n_1} \tau_2^{n_2} \tau_3^{n_3} \tau_4^{n_4}
 \vert
 \ 0\ \leq  p_1\,n_1 +   p_2\,n_2 +   p_3\,n_3 +   p_4\,n_4 \leq n
 \right\}, \quad  p_i,n_i \in \mathbb{N}
\]
which are labeled by certain  characteristic vectors
\[
 (p_1,p_2,p_3,p_4)\,.
\]
For fixed $(p_1,p_2,p_3,p_4)$ such invariant subspaces form a {\it flag}:
\[
 \mathcal{P}_0 \subset \mathcal{P}_1 \subset \ldots \subset \mathcal{P}_k \subset \ldots
\]
where, for simplicity in notations,  we have denoted $ \mathcal{P}^{(p_1,p_2,p_3,p_4)}_n =\mathcal{P}_n$.
The characteristic vector corresponding to the {\it minimal flag} is:
\[
 (1,2,2,3)\,,
\]
which coincides with the minimal flag found  for the rational case.
There are other flags, which are invariant under the action of the Hamiltonian (\ref{hF4algebraic}), are associated with the characteristic vectors $(2,2,3,4)$,
$(2,3,4,6)$, $(2,4,4,6)$ etc. Among these flags there  are two special flags, the first of  which is characterized by the components\footnote{arranged by increasing value, {\it i.e.} following a non-standard Dynkin ordering of the simple roots.} of the Weyl-vector (with respect to simple roots):
\[
 \rho=(8,11,15,21)\,,
\]
and a second flag with characteristic vector defined by the components of the co-Weyl vector :
\[
\rho^{\vee}=(11, 16, 21, 30)\,.
\]

\bigskip

\subsection{Spectrum of the $F_4$-trigonometric model}

In order to find the spectrum of the $F_4$-trigonometric model we should introduce some relevant concepts. \\
The \emph{weight lattice} $L(W)$ is defined as the $\mathbb{Z}$-span of the set
of fundamental weights (\ref{fundweights}) (denoted here as $\{ w_a,a=1\ldots 4\}$). The cone of \emph{dominant weights} $L_{+}(W)$ contains the lattice points
with nonnegative integer coordinates:
\begin{align}
\label{basis}
 L_+ =\left\{\mathbf{n} = \sum_{a=1}^4 n_a w_a = (n_1, n_2, n_3, n_4) ~|\ n_a\geq 0\right\}.
\end{align}

We can define a basis set of monomials
\[
 \left\{
 \tau^{\mathbf n} = \tau_1^{n_1} \tau_2^{n_2} \tau_3^{n_3} \tau_4^{n_4}  \vert   \quad {\mathbf n} \in L_+
 \right\},
\]
which is ordered according to the relation
\[
|{\mathbf n}|^2 > |{\mathbf m}|^2   \Rightarrow   \tau_{\mathbf n} >   \tau_{\mathbf m}
\quad \mbox{for}\quad {\mathbf n}, {\mathbf m} \in L_+ \,.
\]
The Hamiltonian  (\ref{hF4algebraic})  is triangular in this basis. Any eigenfunction can be marked by a dominant
weight $\mathbf{p}$ being  of the form
\begin{equation}
 \label{phip}
 \varphi_{\mathbf{p}} = \tau^{\mathbf p} + \sum_{|\mathbf{m}|^2 < |\mathbf{p}|^2} c_{\mathbf m} \tau^{\mathbf m}
\end{equation}
with energy (see ~\cite{Sasaki:2000}):
\begin{equation}
\label{Ep}
-\epsilon_{\mathbf p}= \beta^2 \bigg((\mathbf{p},\mathbf{p})
+ 2 (\mathbf{p},\varrho) \bigg)\,,
\end{equation}
where $\varrho$ is the deformed Weyl vector (\ref{deformedweyl}).

\subsection{Eigenfunctions}
 We consider the minimal flag of polynomial subspaces $\mathcal{P}^{(1, 2, 2, 3)}_n$, where
\begin{equation}
\label{subspacesminflag}
 \mathcal{P}^{(1, 2, 2, 3)}_n \equiv
 \left\{
 \tau_1^{p_1} \tau_2^{p_2} \tau_3^{p_3} \tau_4^{p_4}
 \vert
 0 \leq  1\,p_1 +   2\,p_2 +   2\,p_3 +   3\,p_4 \leq n
 \right\}, \quad  n = 0,1,2,3\ldots
\end{equation}
The ordering of monomials $\tau^{\mathbf p}$ is given by the norm of the associated dominant weight ${\mathbf p} = (p_1,p_2,p_3,p_4)$, {\it i.e.} by the formula
\begin{eqnarray*}
|\mathbf{p}|^2   &=&
3\,{p_{{3}}}^{2}+8\,p_{{3}}p_{{4}}+6\,{p_{{4}}}^{2}+6\,p_{{2}}p_{{4}}+
2\,{p_{{2}}}^{2}+4\,p_{{2}}p_{{3}}+{p_{{1}}}^{2}+4\,p_{{1}}p_{{4}}+3\,
p_{{1}}p_{{3}}+2\,p_{{1}}p_{{2}}\,.
\end{eqnarray*}
On  the other side, the grading of the monomials $\tau^{\mathbf p}$ is defined as
$1\,p_1 +   2\,p_2 +   2\,p_3 +   3\,p_4$ (for the minimal flag).
Eigenfunctions are labeled by the coordinates of leading dominant weight  in (\ref{phip}) with
an extra super-index labeling the grading $n$ of invariant subspace (\ref{subspacesminflag}) to which the eigenfunction belongs. The eigenfunctions from the spaces ${\mathcal P}_{n}^{({1},{2},{2},{3})},\ n=0,1,2$ are presented in Appendix D.

\bigskip
\section{Conclusion}

We revealed the algebraic-rational nature of the $F_4$-invariant, rational and trigonometric models in the space of orbits.
It is shown that the kinetic energy operator is the Laplace-Beltrami operator with metric with polynomial entries of zero Riemann tensor and the potential is ratio of two polynomials.
We have obtained algebraic forms for the  $F_4$-invariant Hamiltonians, both rational and trigonometric, in terms of the Weyl invariants, polynomial and exponential,
respectively. They are  second order differential operators with  polynomial coefficients.
The both Hamiltonians preserve the same infinite flag of polynomial spaces marked by the characteristic vector $(1,2,2,3)$, indicating exact solvability.
It turns out that for the trigonometric case a solution (formulas) for the Hamiltonian defined
in the standard root space looks easier than one obtained in previous studies~\cite{blt,BLT-trigonometric} where the Hamiltonian was defined in the dual root space.
It allows to unify the description of the $F_4$-trigonometric model under the same systematic description  as it was done for the $G_2,E_{6,7,8}$ trigonometric models (see \cite{BLT-trigonometric}).
Explicit examples of first polynomial eigenfunctions were presented for the rational and trigonometric models. We showed for the first time that both $F_4$-invariant rational and trigonometric models have common particular integral in the space of orbits, which annihilates
the invariant subspace.

\bigskip

This work was supported in part by the University Program FENOMEC, by the PAPIIT
grant {\bf IN109512} and CONACyT grant {\bf 166189}~(Mexico).

\bigskip

\appendix

\section{\it The Weyl-invariant variables $\ta$ for the $F_4$ rational model.}

\begin{eqnarray}
 \ta_2 &=& {x_{1}}^{2} + {x_{2}}^{2} + {x_{3}}^{2} + {x_{4}}^{2}\ , \\
 \ta_6 &=& \frac{1}{6}\left(
 x_{1}^{4} x_{2}^{2} + {x_{1}}^{4}{x_{3}}^{2} + x_{1}^{4}{x_{4}}^{2} + x_{1}^{2} x_{2}^{4}
 - 3\, x_{1}^{2} x_{2}^{2} x_{3}^{2} - 3\, x_{1}^{2} x_{2}^{2} x_{4}^{2}
 \right.\non\\&& \left.
 + x_{1}^{2} x_{3}^{4} - 3\, x_{1}^{2} x_{3}^{2} x_{4}^{2} + x_{1}^{2} x_{4}^{4}
 +  {x_{2}}^{4}{x_{3}}^{2}  +  {x_{2}}^{4}{x_{4}}^{2}  +
{x_{2}}^{2}{x_{3}}^{4}
 \right.\non\\&& \left.
 - 3\, x_{2}^{2} x_{3}^{2} x_{4}^{2} + x_{2}^{2} x_{4}^{4} +
 x_{3}^{4} x_{4}^{2} + x_{3}^{2} x_{4}^{4}
\right)\ ,
  \non\\
 \ta_8 &=& \frac{1}{12}\left(
{x_{1}}^{4}{x_{2}}^{4}-{x_{1}}^{4}{x_{2}}^{2}{x_{3}}^{2}-{x_{1}}^{4}{x_{2}}^{2}{x_{4}}^{2}
 +{x_{1}}^{4}{x_{3}}^{4}-{x_{1}}^{4}{x_{3}}^{2}{x_{4}}^{2}+{x_{1}}^{4}{x_{4}}^{4}-{x_{1}}^{2}{x_{2}}^{4}{x_{3}}^{2}
 \right.\non\\&& \left.
 -{x_{1}}^{2}{x_{2}}^{4}{x_{4}}^{2}
 -{x_{1}}^{2}{x_{2}}^{2}{x_{3}}^{4}+6\,{x_{1}}^{2}{x_{2}}^{2}{x_{3}}^{2}{x_{4}}^{2}
 -{x_{1}}^{2}{x_{2}}^{2}{x_{4}}^{4}-{x_{1}}^{2}{x_{3}}^{4}{x_{4}}^{2}-{x_{1}}^{2}{x_{3}}^{2}{x_{4}}^{4}
  \right.\non\\&& \left.
 +{x_{2}}^{4}{x_{3}}^{4}-{x_{2}}^{4}{x_{3}}^{2}{x_{4}}^{2}+{x_{2}}^{4}{x_{4}}^{4}-{x_{2}}^{2}{x_{3}}^{4}{x_{4}}^{2}
 - {x_{2}}^{2}{x_{3}}^{2}{x_{4}}^{4}+{x_{3}}^{4}{x_{4}}^{4}
\right)\ ,
 \non \\
 \ta_{12} &=& \frac{1}{72}
  \left({x_{1}}^{2}{x_{2}}^{2} + {x_{1}}^{2}{x_{3}}^{2} - 2\,{x_{1}}^{2}{x_{4}}^{2} - 2\,{x_{2}}^{2}{x_{3}}^{2}  + {x_{2}}^{2}{x_{4}}^{2}
           + {x_{3}}^{2}{x_{4}}^{2} \right)
 \non\\&& \hspace{24pt}
\left(
 x_{1}^{2} x_{2}^{2} -2\, x_{1}^{2} x_{3}^{2} + x_{1}^{2} x_{4}^{2} + x_{2}^{2} x_{3}^{2}-2\, x_{2}^{2} x_{4}^{2} + x_{3}^{2} x_{4}^{2}
\right)
\non\\&& \hspace{24pt}
\left(
 2\,x_{1}^{2} x_{2}^{2} - x_{1}^{2} x_{3}^{2} - x_{1}^{2} x_{4}^{2}
 - x_{2}^{2} x_{3}^{2} - x_{2}^{2} x_{4}^{2} + 2\,x_{3}^{2} x_{4}^{2} \right)\ .
 \non
\end{eqnarray}

\section{\it Lowest Eigenfunctions of the $F_4$-rational Model.}

\bigskip

Here the eigenfunctions and eigenvalues corresponding to the spaces ${\mathcal P}_{n}^{({1},{2},{2},{3})}$ for $n=0,1,2$\ are presented in the explicit form:

\begin{itemize}

 \item $n=0$\,  (one eigenstate)
\begin{eqnarray*}
  \phi_0\ &=&\  1 \ ,\\
  \ep_0\  &=&\  0 \ .
\end{eqnarray*}

\item $n=1$\,  (one eigenstate)
\begin{eqnarray*}
  \phi_1\  &=&\ \  \ta_2 - \frac{2}{\om}(6\mu+6\nu+1)\ , \\
  \ep_1\ &=&\  -4\om \ .
\end{eqnarray*}

\end{itemize}

\begin{itemize}

 \item $n=2$\,  (three eigenstates)
\begin{eqnarray*}
 \phi_2^{(1)}\ &=&\ \  \ta_2^2\ -\ \frac{6}{\om}(4\mu+4\nu+1)
\ta_2\ +  \ \frac{6}{\om^2}(4\mu+4\nu+1)(6\mu+6\nu+1)\ , \\
  \ep_2^{(1)}\ &=&\  -8\om\ , \\[5pt]
 \phi_2^{(2)}\ &=&\   \ta_6\ -\ \frac{1}{4\om}(2\mu+4\nu+1)
\ta_2^2\ +\
\frac{3}{4\om^2}(2\mu+4\nu+1)(4\mu+4\nu+1) \ta_2\
\\ &&
  - \frac{1}{2\om^3}(2\mu+4\nu+1)(6\mu+6\nu+1)(4\mu+4\nu+1)\ ,
  \\
 \ep_2^{(2)}\ &=&\  -12\om\ ,\\[5pt]
\phi_2^{(3)}\ &=&\ \  \ta_8 - \frac{1}{\om}(3\nu+1) \ta_6\
+\ \frac{1}{8\om^2}(3\nu+1)(2\mu+4\nu+1) \ta_2^2\
\\&&
 - \frac{1}{4\om^3}(3\nu+1)(2\mu+4\nu+1)(4\mu+4\nu+1) \ta_2\
 \\&&
 +\frac{1}{8\om^4}(3\nu+1)(2\mu+4\nu+1)(6\mu+6\nu+1)(4\mu+4\nu+1)\ ,\\
\epsilon_2^{(3)}\ &=&\  -16\om\ .
\end{eqnarray*}

\end{itemize}

\section{\it The Weyl-invariant variables $\ta$ for the $F_4$ trigonometric model.}

\begin{eqnarray}
 \frac{\ta_1}{2} &=& 8\,
\cos({  \frac{\beta}{2}\,x_{1}}) \cos({  \frac{\beta}{2}\,x_{2}})
\cos({  \frac{\beta}{2}\,x_{3}}) \cos({  \frac{\beta}{2}\,x_{4}})
\non \\
 &&
+
\cos ( \beta\,x_{1} ) + \cos ( \beta\,x_{2} ) +
\cos ( \beta\,x_{3} ) + \cos ( \beta\,x_{4} )
\\[10pt]
\frac{\ta_2}{4} &=&
 \cos ( \beta\,x_{1} ) \cos ( \beta\,x_{2} ) +
 \cos ( \beta\,x_{1} ) \cos ( \beta\,x_{3} )  +
 \cos ( \beta\,x_{1} ) \cos ( \beta\,x_{4} )
\non \\
&&
 +
 \cos ( \beta\,x_{2} ) \cos ( \beta\,x_{3} ) +
 \cos ( \beta\,x_{2} ) \cos ( \beta\,x_{4} ) +
 \cos ( \beta\,x_{3} ) \cos ( \beta\,x_{4} )
\\[10pt]
\frac{\ta_3}{8} &=& 2\,
\cos ( \frac{\beta}{2}\,x_{1} ) \cos ( \frac{\beta}{2}\,x_{2}
)
\cos ( \frac{\beta}{2}\,x_{3} ) \cos ( 3\frac{\beta}{2}\,x_{4}
)
\\
&&
+  2\,
\cos ( \frac{\beta}{2}\,x_{1} ) \cos ( \frac{\beta}{2}\,x_{2}
)
\cos ( 3\frac{\beta}{2}\,x_{3} ) \cos ( \frac{\beta}{2}\,x_{4}
)
\non\\&&
+  2\,
\cos ( \frac{\beta}{2}\,x_{1} ) \cos ( 3\frac{\beta}{2}\,x_{2}
)
\cos ( \frac{\beta}{2}\,x_{3} ) \cos ( \frac{\beta}{2}\,x_{4}
)
\non\\&&
+  2\,
\cos ( 3\frac{\beta}{2}\,x_{1} ) \cos ( \frac{\beta}{2}\,x_{2}
)
\cos ( \frac{\beta}{2}\,x_{3} ) \cos ( \frac{\beta}{2}\,x_{4}
)
\non\\&&
+
\cos ( \beta\,x_{1} ) \cos ( \beta\,x_{2} ) \cos
( \beta\,x_{{3}})  +
\cos ( \beta\,x_{1} ) \cos ( \beta\,x_{2} ) \cos
( \beta\,x_{{4}} )
\non\\&&
+
\cos ( \beta\,x_{1} ) \cos ( \beta\,x_{3} ) \cos( \beta\,x_{4} ) +
\cos ( \beta\,x_{2} ) \cos ( \beta\,x_{3} ) \cos( \beta\,x_{4} )
\non
\\[10pt]
\frac{\ta_4}{16} &=&
\cos^2 ( \beta\,x_{1})\cos (\beta\,x_{2} ) \cos ( \beta\,x_{3} ) +
\cos^2 ( \beta\,x_{1})\cos (\beta\,x_{2} ) \cos ( \beta\,x_{4} )
\\&&
 +
 \cos^2 ( \beta\,x_{1}) \cos (\beta\,x_{3} ) \cos ( \beta\,x_{4}) + \cos ( \beta\,x_{{1}} )
 \cos^2 ( \beta\,x_{2}) \cos( \beta\,x_{3} )
\non\\&&
 +
 \cos ( \beta\,x_{{1}} )
 \cos^2 ( \beta\,x_{2})   \cos ( \beta\,x_{4} )  +
 \cos ( \beta\,x_{{1}} )  \cos ( \beta\,x_{2} )
 \cos^2 ( \beta\,x_{{3}})
\non\\&&
 +
 \cos ( \beta\,x_{1} )    \cos ( \beta\,x_{2} )
 \cos^2 ( \beta\,x_{4}) + \cos ( \beta\,x_{1} )
 \cos^2 ( \beta\,x_{3})   \cos ( \beta\,x_{4} )
\non\\&&
 + \cos ( \beta\,x_{1} ) \cos ( \beta\,x_{3} )
 \cos^2 ( \beta\,x_{4} )  +
 \cos^2 ( \beta\,x_{2} )
 \cos (\beta\,x_{3} ) \cos ( \beta\,x_{4} )
\non\\&&
 +
 \cos( \beta\,x_{2} )  \cos^2 (\beta\,x_{3})
 \cos ( \beta\,x_{4} ) +
 \cos( \beta\,x_{2} ) \cos ( \beta\,x_{3} )\cos^2 ( \beta\,x_{4} )
\non\\&&
 -
 \cos ( \beta\,x_{1} ) \cos ( \beta\,x_{2} )
-\cos ( \beta\,x_{1} ) \cos ( \beta\,x_{3} )
-\cos ( \beta\,x_{1} ) \cos ( \beta\,x_{4} )
\non\\&&
 -\cos ( \beta\,x_{2} ) \cos ( \beta\,x_{3} )
 -\cos ( \beta\,x_{2} ) \cos ( \beta\,x_{4} )
 -\cos ( \beta\,x_{3} ) \cos ( \beta\,x_{4} )
 \non
\end{eqnarray}

\bigskip

\section{\it Lowest Eigenfunctions of the $F_4$-trigonometric Model.}

\bigskip

Here, the eigenfunctions and eigenvalues corresponding to the spaces ${\mathcal P}_{n}^{({1},{2},{2},{3})}$ for $n=0,1,2$\ are presented in the explicit form:

\bigskip

$\bullet \quad n=0$\,. Ground state:

\begin{eqnarray*}
\varphi_{[0,0,0,0]}^{(0)} &=& 1 \,,  \\
\varepsilon_{[0,0,0,0]}^{(0)} &=& 0\,.
\end{eqnarray*}

$\bullet \quad n=1$\,. One eigenstate:

\[
 \mbox{monomials: } [1,\tau_1],\quad   |\mathbf{p}|^2:  [0, 1],\quad \mbox{gradings: } [0, 1]
\]

\begin{eqnarray*}
\varphi_{[1,0,0,0]}^{(1)} &=& 1 + \frac{1}{24}\,{\frac { \left( 1+6\,\nu+5\,\mu \right)  }{\mu}} \tau_{{1}}\,,
\\
\varepsilon_{[1,0,0,0]}^{(1)} &=& -(1+6\,\nu+5\,\mu)\,,
\end{eqnarray*}


\bigskip

$\bullet \quad n=2$\,. Three eigenstates:

\bigskip

\(
(i) \quad \mbox{monomials: } [1, \tau_1, \tau_2], \quad  |\mathbf{p}|^2: [0, 1, 2], \quad \mbox{ gradings:} [0, 1, 2]
\)

\begin{eqnarray*}
\varphi_{[0,1,0,0]}^{(2)} &=&
24\,  {\frac { \left( 3\,{\mu}^{2}+\mu\,\nu+4\,{\nu}^{2}+\nu \right) }{ \left( 4\,\nu+\mu+1 \right)  \left( 1+5\,\nu+3\,\mu \right) }}
 + 6\,{\frac {\mu\, }{4\,\nu+\mu+1}} \tau_{{1}} + \tau_{{2}}\,,
\\
\varepsilon_{[0,1,0,0]}^{(2)} &=&   -2(1 + 5 \,\nu +3\,\mu)\,.
\end{eqnarray*}

\bigskip

\(
(ii) \quad  \mbox{monomials: } [1, \tau_1, \tau_2, \tau_3 ],\quad   |\mathbf{p}|^2: [0, 1, 2, 3], \quad \mbox{ gradings:} [0, 1, 2, 2]
\)

 \begin{eqnarray*}
 \varphi_{[0,0,1,0]}^{(2)} &=&
 8\,{\frac { \left( 6\,{\mu}^{2}+9\,\mu\,\nu+\mu+8\,{\nu}^{2}+3\,\nu
 \right)  }{ \left( 3\,\nu+2\,\mu+1 \right)  \left( 1+4\,\nu+3\,\mu \right) }}
+ {\frac {  \left( 6\,{\mu}^{2}+5\,\mu\,\nu+\mu+2\,{\nu}^{2}+\nu \right) }
{\mu\, \left( 3 \,\nu+2\,\mu+1 \right) }}\tau_{{1}}
\\&&
+ \tau_{{2}}  + \frac{1}{12}\,{\frac { \left( 3\,\mu+1+2\,\nu \right)}{\mu}}  \tau_{{3}}\,,
\\
\varepsilon_{[0,0,1,0]}^{(2)} &=&   -3(1 + 4\nu +3\mu)\,.
\end{eqnarray*}

\bigskip
\bigskip

\(
(iii) \quad \mbox{monomials: }
 [1, \tau_1, \tau_2, \tau_3, \tau_1^2], \quad  |\mathbf{p}|^2:  [0, 1, 2, 3,4], \quad \mbox{ gradings:}  [0, 1, 2, 2, 2]
\)


\begin{eqnarray*}
\varphi_{[2,0,0,0]}^{(2)} &=&
-8\,{\frac { \left( 8\,{\mu}^{3}-15\,{\mu}^{2}+4\,{\mu}^{2}\nu-32\,\mu\,\nu-11\,\mu-3-24\,{\nu}^{2}-18\,\nu \right) }
{ \left( 5\,\mu+3+6\,\nu \right)  \left( 2+6\,\nu+5\,\mu \right) }}
\\&&
-2\,{\frac { \left( 8\,{\mu}^{3}-9\,{\mu}^{2}+4\,{\mu}^{2}\nu-
10\,\mu\,\nu-9\,\mu-6\,\nu-2-4\,{\nu}^{2} \right) }{ \left(
5\,\mu+3+6\,\nu \right)  \left( 1+3\,\mu \right) }}\tau_{{1}}  + \tau_{{2}}
\\&&
+\frac{1}{3}\,{\frac { \left( 2\,\mu+1+\nu \right) }{1+3\,\mu}} \tau_{{3}}
-\frac{1}{6}\,{\frac { \left( 2\,{\mu}^{2}+\mu\,\nu+3\,\mu+1+\nu \right)}
{1+3\,\mu}} {\tau_{{1}}}^{2}\,,
\\
\varepsilon_{[2,0,0,0]}^{(2)} &=& -2(2+6\,\nu+5\,\mu)\,.
\end{eqnarray*}

\appendix


\end{document}